# Impact of Financial Inclusion on the Socio-Economic Status of Rural and Urban Households of Vulnerable Sections in Karnataka

*Dr. Manohar V. Serrao
**Dr. A. H. Sequeira
***Dr K.V.M Varambally

* Associate Professor, Dept of Economics, St Aloysius Evening College, Mangalore, India, Email: *manoharvserrao@gmail.com*

** Professor, School of Management, National Institute of Technology Karnataka Surathkal, India, Email: *aloysiushs@gmail.com*

*** Professor, Manipal Institute of Management, Manipal University, India,
Email: kv.varambally@manipal.edu

## Abstract:

*'Financial inclusion' and 'inclusive growth' is the buzz words today. Inclusive growth empowers people belonging to vulnerable sections. This in turn depends upon a variety of factors- the most important being "financial inclusion", which plays a strategic role in promoting inclusive growth and helps in reducing poverty by providing regular and reliable sources of finance to the vulnerable sections. In this direction, the Government of India in its drive for financial inclusion has taken several measures to increase the access to and availing of formal financial services by unbanked households. The purpose of this paper is to assess the nature and extent of financial inclusion and its impact on the socio-economic status of households belonging to vulnerable sections focusing on inclusive growth. This has been analyzed with the theoretical background on financial access and economic growth, and by analyzing the primary data collected from the Revenue Divisions of Karnataka. The results show that there is a disparity in nature and extent of financial inclusion. Access to, availing of formal banking services pave the way to positive changes in the socio-economic status of households belonging to vulnerable*

*sections which are correlated, leading to inclusive growth based on which paper proposes a model to make the financial system more inclusive and pro-poor.*

# 1. Introduction

In economic literature 'inclusive growth' is defined as an equitable allocation of resources, where benefits accrue to all the sections of the society. It is the involvement of all sections and regions of society in the growth of the economy and getting the benefits of growth, which achieves the equity objective in growth. The concept of inclusive growth is referred to as 'economic growth with equal opportunities to all. It is nothing but creating growth opportunities and making them accessible to all particularly to the poor (Ali & Zhuang, 2007).

Financial inclusion is the delivery of financial services at an affordable cost to the vast sections of the disadvantaged and low-income groups. The Rangarajan Committee, Government of India (2008), defines financial inclusion as 'the process of ensuring access to financial services and timely, adequate credit where needed, to vulnerable groups such as weaker sections and low-income groups, at an affordable cost'. This is an effort of 'mainstreaming the marginalized' which plays a key role in the process of inclusive growth involving all the sections of the population and regions of the economy. Therefore, access to finance is a pre-condition for poverty eradication is an integral part of the growth process. But the in-ground reality the households of vulnerable sections are denied access to formal financial services. This is popularly known as financial exclusion, which occurs due to inaccessibility, distances, inadequate infrastructure, and low absorptive capacity of the poor households that hinders growth. Data on indicators of access to finance at all Indian level, and in Karnataka, confirm a similar scenario. To study this problem further and to have an effective pro-poor financial inclusion policy that achieves the equity objective through poverty reduction, equal distribution of resources and capacities is the need of the hour. This challenge of promoting inclusive growth by providing access to formal financial services is the background of this paper.

In India, the role of finance in promoting equitable growth has been emphasized since the beginning of national plans. Some of the important steps initiated in this direction are the establishment of rural cooperatives, Regional Rural Bank's (RRB), National Bank for Agriculture and Rural Development (NABARD), nationalization of commercial banks, focus given in government policies on social banking, rural credit, priority sector lending, Lead Bank Schemes, interest rate ceilings, subsidies, etc. Even after this policy push in India towards financial inclusion, formal banking services have not penetrated the vast segments of households and the poor and marginal sections of the society do not have access to any financial services.

## 2. Dimensions of Financial Inclusion

An exploration of the supply and the demand side of financial inclusion reveals that internationally there is a wider perspective regarding the concept of financial inclusion. In India, presently, financial inclusion is confined to providing minimum access to a savings bank account to all. It is, having minimum facilities of banking by the general public or citizens of the country. Along with this, there are two extreme possibilities of access to finance. First, with the group of customers who get access to all the formal financial services and second with the group which is financially excluded and denied access to the most basic financial products. In between these two possibilities, there are customers of banks who use the banking services only for deposits and withdrawals without the flexibility of access to finance. This scenario of multiple possibilities indicates that financial inclusion is a multi-faceted concept with several dimensions relevant to the specific region's agenda.

## 3. Role of Finance in Economic Development

In the literature on the role of finance in economic development, it is advocated that financial development creates pro-growth conditions in an economy, through the demand-supply mechanism. In the eighteenth century, Adam Smith (1776) had expressed the view that there is a significant relationship between the high density of banks in Scotland and the development of the Scottish economy. Walter Bagehot (1873) and John Hicks (1969) argued that financial inclusion played a determining role in the industrialization of England, through capital mobilization. Schumpeter (1912) had the view that a well-functioning banking network leads to technological innovations. In the year 1952, Joan Robinson opined that there is a mutual relationship between 'economic development and demand for financial services and the economic system responds positively to this change'. Simon Kuznets (1955, 1963) and Kaldor (1966) pointed out a trade-off between growth and social justice in the early stages of development until the benefits of growth spread throughout the economy.

In the literature of the World Bank (2005) on the modern development theory, it is stated that 'the progression of financial access, growth, and income dynamics of different generations are closely related.

A group of social scientists (Aghion & Bolton, 1997; Aghion, Caroli & Garcia-Penalosa, 1999; Banerjee &Newman, 1993; Galor & Zeria1993; Rajan & Zingales, 2003), argued that, given the capital market imperfection, vulnerable section households who have high marginal productivity, will have little money to invest in their education and their occupational options are limited because low initial endowments. King and Levine (1993) pointed out that, banking access is an important incentive for technological innovation. A well-developed financial system paves the way to faster and equitable growth. An index to measure access to finance by Patrick Honohan in 2004 in 160 countries revealed that economies with higher indices were advanced economies and societies with the deeper financial system had a low level of absolute poverty.

Pitt and Khandker's Grameen Bank and MFI study in Bangladesh in the year 1998 proved a significant and positive effect of the use of credit on household expenditures; assets, labor supply, and the possibility of children attending schools. Coleman's study on microcredit borrowers in northeast Thailand in the year 1999 proved an insignificant impact of credit on physical assets, savings, expenditure on health care, education, etc.

Gine and Townsend (2004), in their study between 1976 and 1996 on Thai households reveal that flexibility in financial access leads to an increase in access to credit services and explains the quick growth in per capita GDP in the economy.

Burgess and Pande (2005), in one of the experiments on the Indian government's bank branch expansion policy, between 1977 and 1990, found that there was a fast growth in non-agricultural output and decline in poverty after the branching regulations injected and the new bank branches opened.

Kempson, (2006) points out the literature highlighting a positive correlation between the Index of Financial Inclusion and the ranks of UNDP Human Development Index (HDI) highlighting the countries of high-income inequality with less formal financial access.

Demirguc Kunt and Levine (2007) argued that reducing financial market imperfections acts as a positive incentive resulting in the expansion of individual opportunities.

World Bank research report on access to finance (2008), states that financial access can have direct and indirect benefits on small firms and poor households, makes them more capable to take advantage of investment opportunities, and insures them against risks. Studies by using cross-sectional data on households in Peru revealed that after availing formal finance there were positive changes in household consumption expenditure, income prospects, and the number of children attending school. Therefore, without an inclusive financial system, poor individuals and small enterprises have to rely on their limited savings and earnings to invest in their economic and educational activities and to take the advantage of growth opportunities.

Michael Thiel (2001) opined that, even though there is a firm consensus in the economic literature underscoring the role of a well-functioning financial sector in the efficient allocation of resources and the exploitation of growth potential, the economic literature is also less consensual on how and to what extent finance affects economic growth. In this direction present paper is an attempt to consolidate the empirical proof of the impact of financial inclusion on the socio-economic status of vulnerable section households of rural and urban segments in Karnataka with the following objectives.

## 4. Objectives

1. To analyze the nature and extent of financial inclusion among the households of vulnerable sections across the rural and urban segments.

2. To assess the economic and social impact of financial inclusion on households belonging to the vulnerable sections.

3. To suggest a model for the effective financial inclusion policy that would benefit households.

## 5. The Conceptual Framework

To meet the objectives of analyzing financial inclusion and its impact on households of vulnerable sections a conceptual framework linking financial inclusion, the impact of availing financial services and inclusive growth of the economy has been arrived at based on the theoretical background (Figure 1). Accordingly, the nature of financial inclusion has been analyzed by looking at the type and number of formal financial institutions accessed and services availed by the households. The extent of financial inclusion has been quantified by looking at the outreach of financial services on the various income and asset levels and the number of years of availing formal financial services. The socio-economic impact of financial inclusion has been assessed and quantified by exploring the material, cognitive, perceptional, and relational changes which have been occurred in the household after availing formal financial services.

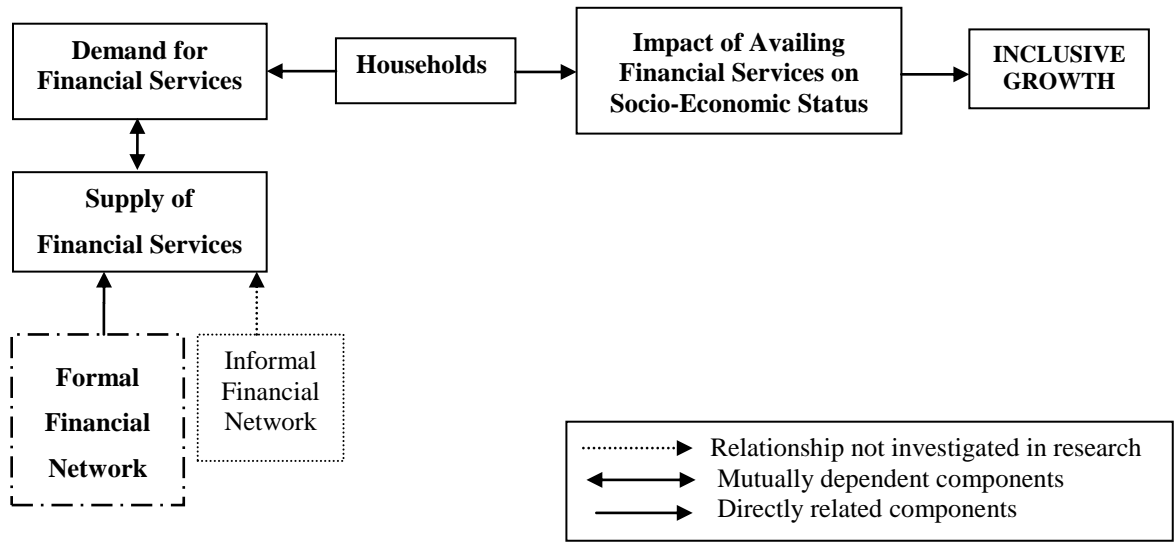

*Source: Literature Review*

**Figure 1: Conceptual Framework to Investigate Nature, Extent, and Impact of Financial Inclusion**

## 6. Sources of Data and Methodology

Data was gathered both from primary and secondary sources. Primary data was collected from the sample area, through the fieldwork, by using the sample survey method. To select the units of population multi-stage sampling method was used. In this regard, direct interviews of the heads of the rural and urban households of vulnerable sections across the revenue divisions of Karnataka were conducted by using questionnaire schedules.

Primary data was collected by using the cross-sectional method, to get a snapshot of the variables of interest at a single point in time. It has been analyzed to draw inferences on the nature, extent, and impact of financial inclusion on the socio-economic status of rural and urban households. To measure the characteristics of the units of population, both qualitative and quantitative methods of measurement were used.

Data were summarized and assessed by frequency and percentages for categorical data. To analyze the interval data and for rating scales Mean, Standard Deviation, and Median was used. Multiple regression analysis has been used to evaluate the economic and social impact of financial inclusion on households belonging to the rural and urban segments.

In the study, while measuring the components of financial inclusion and its impact on the socio-economic status, the total score of the individual sub-components of the constructs was obtained by adding up the scores of every question coming under those components followed by standardization and sorted out using the property of normal distribution. The component of financial inclusion and its impact was measured as an independent component and the final component of inclusive growth was considered as a dependent component. They were assessed and categorized as average, below average, and above-average using the properties of normal distribution and percentile values.

## 7. Assumptions

The basic assumption of the study is that other things being constant, financial inclusion leads to positive changes in the socio-economic conditions of the households, resulting in the 'inclusive growth' of the households of rural and urban segments. In the present study, 'inclusive growth', is defined as the 'growth' or the 'positive changes' taken place in the socio-economic status of the rural and urban households after availing formal financial services, measured on the combined scores of the components of 'financial inclusion and the 'impact of availing formal financial services.

## 8. Present Status of Financial Inclusion in India

According to the 68$^{th}$ National Sample Survey report, Government of India, over 80 million poor people are living in the cities and towns of India and they lack access to the most basic banking services such as savings accounts, credit, remittances, and payment services, financial advisory services, etc., and they depend on the informal sector for their savings and loan requirements. The financially excluded sections comprise largely rural masses comprising of marginal farmers, landless laborers, oral lessees, self-employed and unorganized sector enterprises, urban slum dwellers, migrants, ethnic minorities and socially excluded groups, senior citizens, and women.

Census data 2011, GOI, reveals that, at all India level, in general, 58.7 percent of households availed banking services. Across the states, in Andaman and Nicobar

Islands, a maximum of about 89.3 percent of households were availing formal banking services, followed by Haryana with 89.1 percent. In Nagaland, only 34.9 percent of households had accessed banking services. In Karnataka, the percentage of households availing banking services was above average with 61.1 percent.

Credit Delivery and Financial Inclusion report of RBI, Government of India (2012) states, out of 199 million households in India, only 68.2 million households had access to banking services. As far as rural areas were concerned, out of 138.3 million rural households in India, only 41.6 million rural households had access to basic banking services. In respect of urban areas, only 49.52 percent of urban households had access to banking services and only 34 percent of India's urban population with annual income less than 50,000 had access to banking services.

## 9. Supply side of Financial Inclusion in the State of Karnataka

Financial Inclusion was initiated in the state of Karnataka by the regional office of RBI, in December 2005. Rural Planning and Credit Department (RPCD) requested the State Level Bankers Committee (SLBC), Karnataka, to launch the first phase of financial inclusion in the state.

According to the SLBC report (2012), to provide formal financial services in Karnataka, there were 41 Commercial Banks (CB) with a network of 5122 branches. Along with this, there were other institutions like Karnataka State Cooperative Apex Bank (KSCAB) with 31 branches, nearly 21 District Credit Cooperative Banks (DCCB) with 615 branches and 4613 Primary Agriculture Cooperative Societies (PACS), about 6 Grameena Banks with 1256 branches, The Karnataka State Cooperative Agriculture and Rural Development Bank (KSCARDB) with 177 Primary Cooperative Agriculture and Rural Development Bank branches (PCARDB). Besides, Urban Cooperative Banks & other Cooperative institutions were also providing banking services in the State. According to the report, the number of people served by each bank branch in Karnataka was 8500, excluding PACS, and 5106 including PACS.

An exploration of the secondary data regarding the universal access to formal banking services in the sample area across the revenue divisions reveals that formal banking access is provided universally, without any discrimination, to all (Table 1).

**Table 1: Formal Financial Network Units Providing Access to Financial Services in the Sample Area**

| Taluks | Nationalized Banks | RRB | DCC Banks | KASKARD Banks | Co-operative Societies/Units |
|---|---|---|---|---|---|
| **Mangalore** | 283 | 08 | 15 | 02 | 207 |
| **Anekal** | 44 | 04 | 01 | 02 | 141 |
| **Bijapur** | 43 | 19 | 08 | 01 | 519 |
| **Basavkalyana** | 14 | 11 | 07 | 01 | 347 |

*Source: Districts Statistics, Official Website of Government of Karnataka 2013*

Even though Karnataka is the cradle of the banking industry and there is universal access to the formal banking services, an exploration of secondary data on the percentage of households availing banking services across the revenue divisions, according to the 2011 census report reveals that about 59 percent of the rural households and nearly 64.34 percent of urban households accessed formal banking network to avail financial services. Across the revenue divisions, in the Udupi district, nearly 93 percent of households accessed formal banking network, followed by Dakshina Kannada district with 87 percent which belongs to the Mysore revenue division. In both the districts, the maximum percentage of households accessed formal banking services compared with the districts of other revenue divisions.

Among the districts of the Guburga revenue division, Yadgir bottoms the table in Karnataka state in which nearly 39 percent of the households accessing the formal banking network.

Therefore, an examination of the secondary data on the access to and availing of formal banking services at all India level and in the state of Karnataka reveal that, even though there is universal access, there is a disparity in financial inclusion at the macro level of the Indian economy, and formal banking services are not evenly distributed in the state.

# 10. Observations on Nature, Extent, and Impact of Financial Inclusion among the Rural and Urban Segments.

To study the nature, extent, and impact of financial inclusion among the rural and urban segments across the revenue divisions of Karnataka, nearly 1051 Below Poverty Line (BPL) households were selected. Out of this, nearly 504 were rural category households and 547 urban category BPL households. Across the revenue divisions, nearly 28.3 percent of urban households were from Anekal taluk in the Bangalore revenue division, about 24 percent each from Bijapur taluk of Belgaum revenue division and Basavakalyana taluk of Gulbarga revenue division. Around 23.8 percent of urban households selected were from Mangalore taluk of Mysore revenue division. Thus, about 48 percent of households selected were of rural category and 52 percent urban category households (Figure 2).

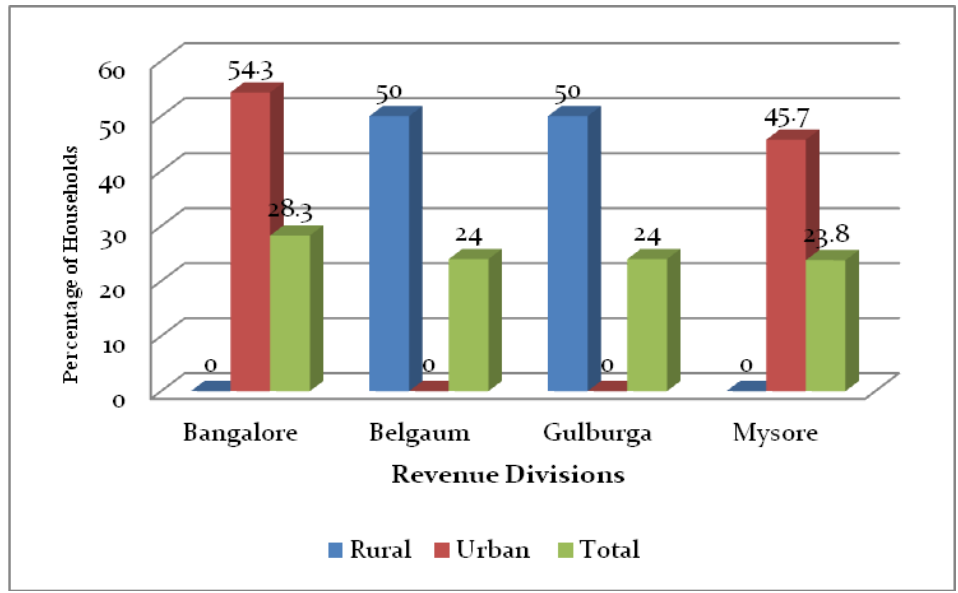

*Source: Survey Data*

**Figure 2: Category of Households across Revenue Divisions**

Regarding the distance to avail banking services, nearly 78.6 percent of the households had the distance of fewer than 5 kilometers to access basic or other banking facilities, about 20.6 percent of the households were within 5 to10 kilometer range of formal banking outlets and only 0.8 percent households had to cover above 10 kilometers to avail banking services. Rural BPL households had a distance problem to avail basic banking and other facilities when compared to their urban counterparts, and it was statistically highly significant, confirmed by the Chi-Square test ($x^2$= 117.353, df=1, p=.000<0.01, HS) (Table 2).

**Table 2: Distance as Barrier in Accessing Banking Services**

(Figures in percentage)

| Distance as Barrier in Availing Banking Services | Region | | Total |
|---|---|---|---|
| | Rural | Urban | |
| **No** | 73.8 | 97.1 | 85.9 |
| **Yes** | 26.2 | 2.9 | 14.1 |
| **Total** | 100.0 | 100.0 | 100.0 |

*Note: $x^2$ =117.353, d=1, p =0.000<0.01, HS*

*Source: Survey Data*

Among the 1051 rural and urban households surveyed across the revenue divisions of Karnataka, CB's emerged as the major channel for providing banking services. The majority of the households were accessing two and more than two formal financial institutions to avail themselves the banking services, were the beneficiaries of poverty eradication, risk mitigation, productivity enhancement, and human development programs sponsored by the Government.

Across the revenue divisions, there was a disparity in access to and availing of formal banking services which were not penetrated the lower-income category, which started availing banking services in the recent past. In this regard, it was observed that, among 65.5 percent of the rural households, the nature of financial inclusion was below-average level.

Only 6.9 percent of rural households were in the above-average category and nearly 47.9 percent of the urban households had an above-average level nature of financial inclusion, while 9.1 percent were in the category of below average. In aggregate, about 36.2 percent of households had a nature of financial inclusion below average and 28.3 percent of households were in the above-average category. Across the revenue divisions, the difference between rural and urban households was statistically significant, confirmed by the Chi-Square test ($x^2$= 403.372, df=2, p=0.000<0.01 HS) (Table 3).

**Table 3: Distribution of Households based on the Nature of Financial Inclusion**

(Figures in percentage)

| Nature of Financial Inclusion | Region | | |
|---|---|---|---|
| | Rural | Urban | Total |
| **Below Average** | 65.5 | 9.1 | 36.2 |
| **Average** | 27.6 | 43.0 | 35.6 |
| **Above Average** | 6.9 | 47.9 | 28.3 |
| **Total** | 100.0 | 100.0 | 100.0 |

Note: $x^2$=403.372, df=2, p=0.000<0.01, HS

*Source: Survey Data*

On average, in 79.2 percent of rural households, the extent of financial inclusion was below-average level. In aggregate, about 7.6 percent of households had the extent of financial inclusion above-average level. Among the urban households, nearly 31.2 percent of households had the extent of financial inclusion below average and around 56.5 percentage households had it above average level. The difference between rural and urban households across the revenue divisions was statistically highly significant, revealed by the Chi-Square test ($x^2$= 245.720, df=2, p=0.000<0.01HS) (Table 4).

**Table 4: Distribution of Households based on the Extent of Financial Inclusion**

(Figures in percentage)

| The extent of Financial Inclusion | Region | | |
|---|---|---|---|
| | Rural | Urban | Total |
| Below Average | 79.2 | 31.2 | 56.1 |
| Average | 3.3 | 12.3 | 7.6 |
| Above Average | 17.6 | 56.5 | 36.3 |
| Total | 100.0 | 100.0 | 100.0 |

Note $x^2$: =245.720, df=2, p=0.000<0.01, HS

*Source: Survey Data*

In total, across the revenue divisions, around 51 percent of rural households had financial inclusion below average, and in 20.2 percent households, it was above average level. In the urban category, analysis of data revealed that nearly 29.1 percent of households were in the below-average level category of financial inclusion and about 24.7 percent of households were in the above-average category. The difference between the rural and urban households across the revenue divisions was statistically significant, confirmed by the Chi-Square test ($x^2$= 55.321, df=2, p=0.000<0.01 HS) (Table 5).

**Table 5: Distribution of Households based on the Level of Financial Inclusion**

(Figures in percentage)

| Level of Financial Inclusion | Region | | |
|---|---|---|---|
| | Rural | Urban | Total |
| Below Average | 51.0 | 29.1 | 39.6 |
| Average | 28.8 | 46.3 | 37.9 |
| Above Average | 20.2 | 24.7 | 22.5 |
| Total | 100.0 | 100.0 | 100.0 |

*Note:* $x^2$=55.321, df=2, p=0.000<0.01, HS

*Source: Survey Data*

After availing the formal banking services there were changes occurred in the socio-economic conditions of the households' viz., household assets, income, property ownership, consumption expenditure and the number of crops cultivated, housing conditions, livelihood skills, confidence, mobility, and self-esteem level of the members, empowerment of women, etc., which were quantified in terms of material, cognitive, perceptional and relational changes occurred after availing formal banking services.

Accordingly across the revenue divisions, after availing formal banking services, in 26.0 percent of households material changes were below average, about 60.2 percent of households experienced average level material changes and nearly 13.8 percent of households were in the above-average category of material changes. In the Mysore revenue division, about 75.2 percent of households were in the category of average level material changes, which was the highest across the revenue divisions, whereas, in 47.6 percent of households of the Belgaum division, the material changes were below average followed by the Gulbarga revenue division with 29.8 percent.

The difference across the revenue divisions in thisregard was statistically significant as confirmed by the Chi Square test ($x^2$=108.986, df=2, p=0.000<0.01 HS) (Figure 3).

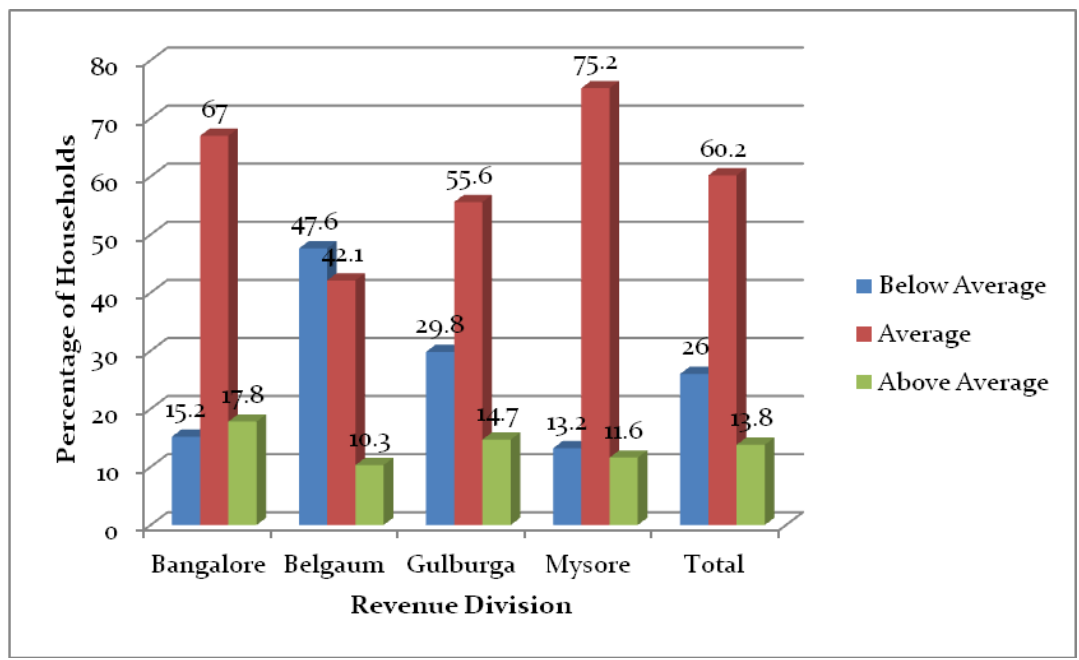

Note: $x^2$ =108.986, df=2, p=0.000<0.01, HS
*Source: Survey Data*

**Figure 3: Level of Material Changes after Availing Financial Services**

In the category of the cognitive change, across the revenue divisions, in 9.3 percent of households, cognitive changes were below average level, around 56.0 percent of households experienced average level cognitive changes and nearly 34.6 percent of households were in the above-average category of cognitive changes. In the Gulbarga revenue division, nearly 79.4 percent of households were in the category of average level cognitive changes, which was the highest across the revenue divisions followed by nearly 52.4 percent of households experiencing cognitive changes in the Belgaum division. In 45.1 percent of households of the Bangalore revenue division, cognitive changes were above average level. Around 12.7 percent of households of Gulbarga revenue division, the cognitive changes were below average level. The difference in this regard across the revenue divisions was statistically significant as confirmed by the Chi-Square test ($x^2$=107.145, df=6, p=0.000<0.01 HS) (Figure 4).

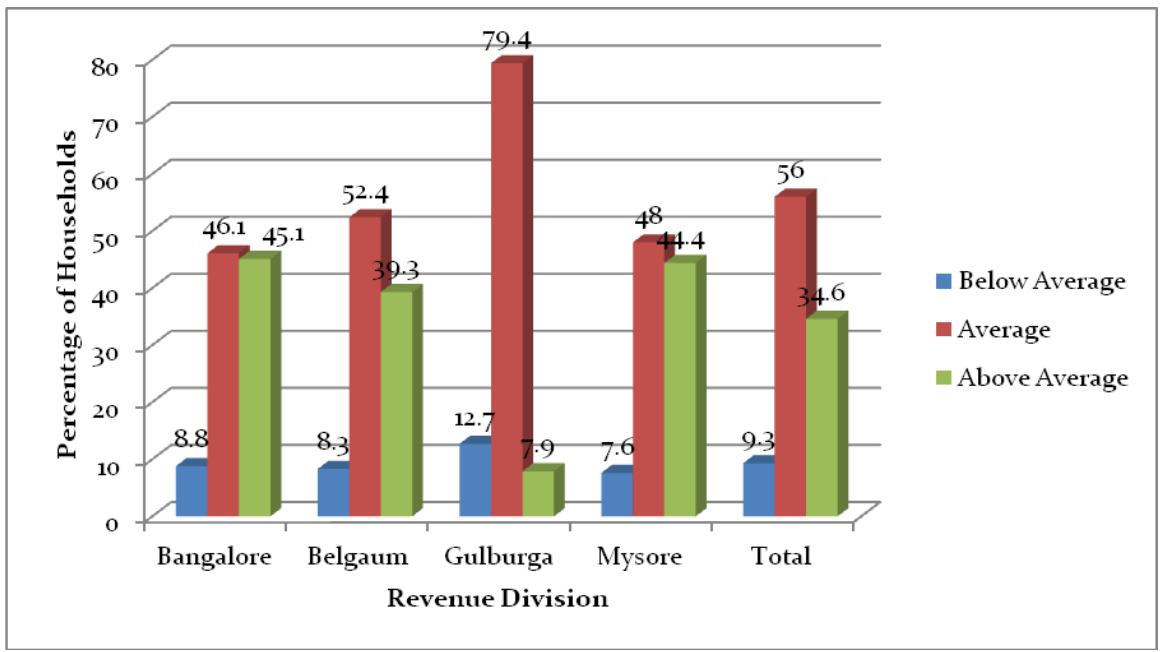

Note: $x^2$ =107.145, df=6, p=0.000<0.01 HS

*Source: Survey Data*

**Figure 4: Level of Cognitive Changes after Availing Financial Services**

Across the revenue divisions, in 47.5 percent of households, perceptional changes were below average, around 1.6 percent of households experienced average level changes and nearly 50.9 percent of households were in the above-average category of perceptional changes after availing financial services. In the Mysore revenue division, almost 92.8 percent of households were in the category of above-average level perceptional changes, which was the highest across the revenue divisions, whereas, in 89.3 percent of households of the Belgaum division, perceptional changes were below average. The difference across the revenue divisions was statistically significant as confirmed by Fisher's Exact test (p=0.000<0.01 HS) (Figure 5).

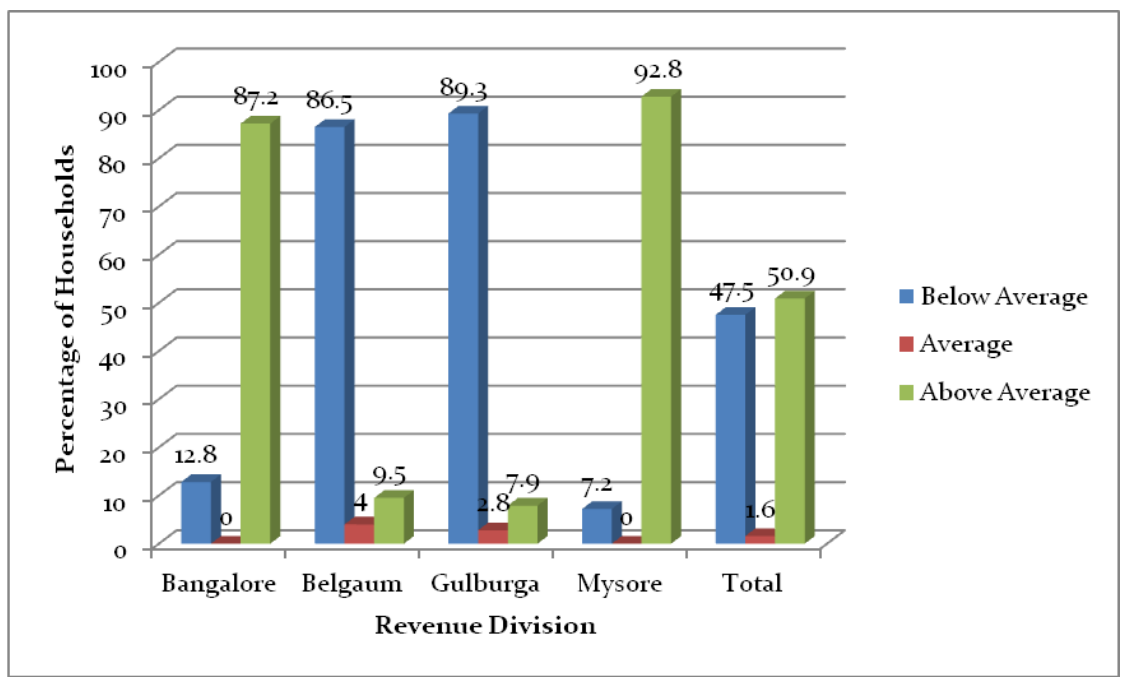

*Note:* Fisher's Exact Test, p=0.000<0.01, HS
*Source: Survey Data*

**Figure 5: Level of Perceptional Changes after Availing Financial Services**

Across the revenue divisions, in 0.7 percent of households, relational changes were below average, around 88.1 percent of households experienced average level relational changes and nearly 11.2 percent were in the above-average category. In the Belgaum revenue division, about 96.4 percent of households were in the category of average level relational changes, which was the highest across the revenue divisions, followed by nearly 90.8 percent of households of the Mysore division. In 23.8 percent of households of the Gulbarga revenue division, relational changes were above average. The difference across the revenue divisions was statistically significant as confirmed by Fisher's Exact test (p=0.000<0.01HS) (Figure 6).

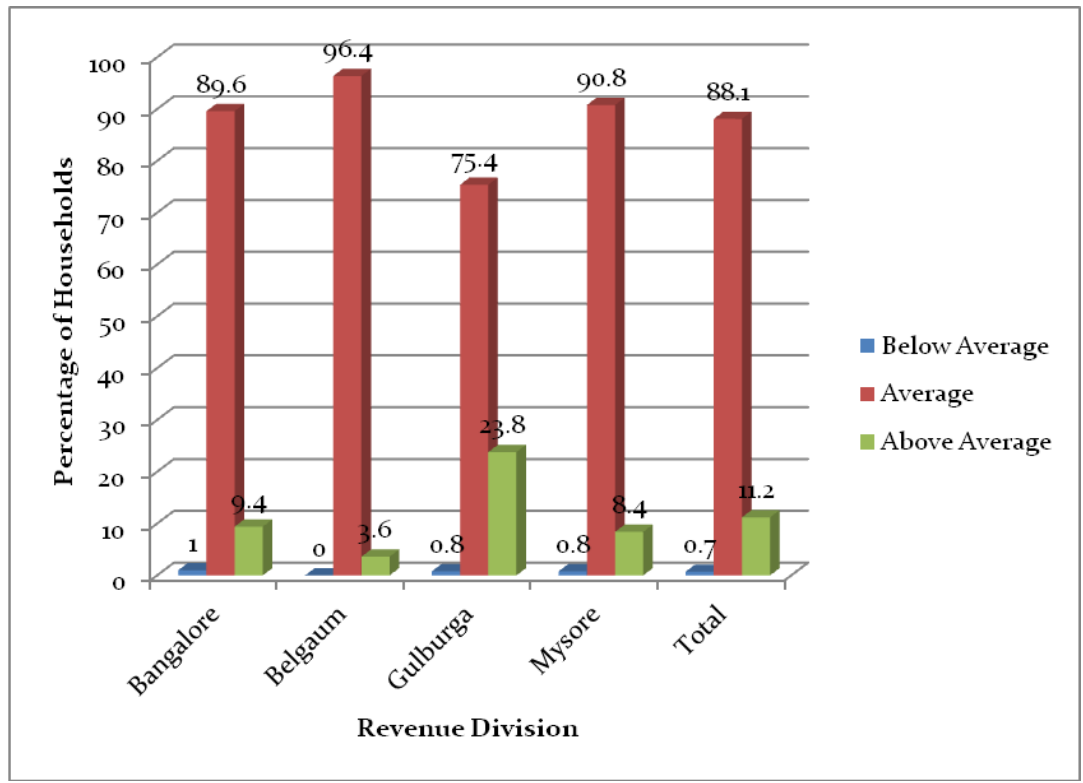

Note: Fisher's Exact Test, p=0.000<0.01, HS
*Source: Survey Data*

**Figure 6: Level of Relational Changes after Availing Financial Services**

By adding up the socio-economic components of the impact of availing formal financial services, the total impact of availing formal financial services was measured. Accordingly, in 23.3 percent of households, the impact of availing formal financial services was below average, around 46.3 percent of households experienced the average level impact of availing financial services and in 30.4 percent of households, it was above average level. Nearly 58.3 percent of households in the Gulbarga division had the average level impact of availing financial services, whereas in 44.4 percent households of Belgaum division it was below average. The difference across the revenue divisions was statistically significant as confirmed by the Chi-Square test ($x^2$=252.915, df=2, p=0.000<0.01 HS) (Figure 7).

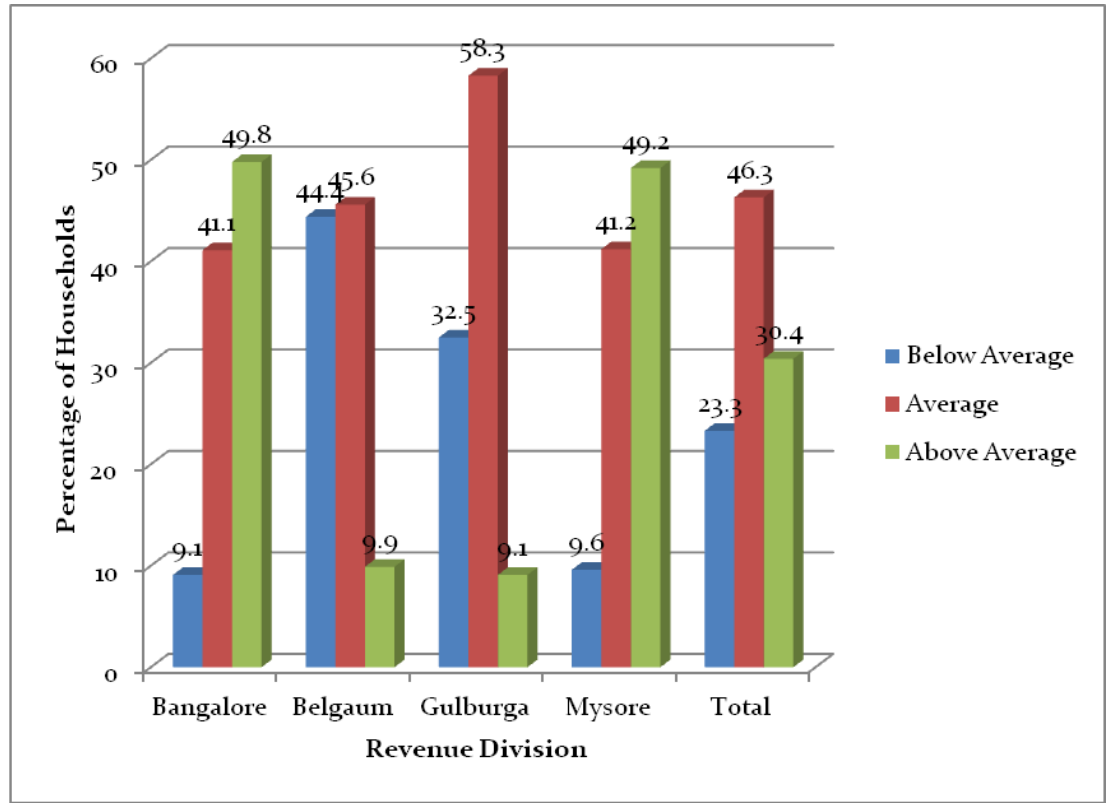

Note: $x^2$=252.915, df=2, p=0.000<0.01, HS

*Source: Survey Data*

**Figure 7: Impact of Availing Financial Services in Aggregate**

By adding up the components of financial inclusion and the impact of availing banking services, the outcome component, the 'inclusive growth' was measured. Accordingly, in 31.0 percent of households, the inclusive growth was below average level, around 40.6 households had average level inclusive growth and in 28.4 percent households, inclusive growth was above average level. About 60.0 percent of the households in the Mysore division had above-average level inclusive growth, whereas, around 69 percent of households in the Gulbarga revenue division had below-average level inclusive growth after availing formal financial services. The difference in inclusive growth across the revenue divisions was statistically significant as confirmed by the Chi-Square test ($x^2$=377.227, df=6, p=0.000<0.01 HS) (Figure 8).

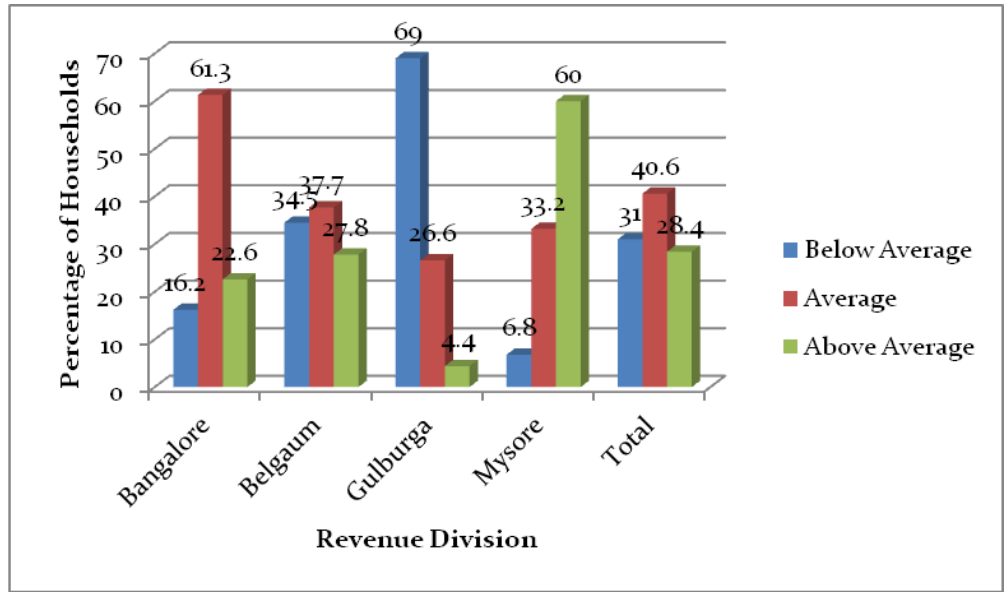

Note: $x^2$ =377.227, df=6, p=0.000<0.01, HS

*Source: Survey Data*

**Figure: 8 Level of Inclusive Growth after Availing Financial Services**

## 11. Correlation Analysis

Correlation between the components of the constructs of the study was assessed by Karl Pearson's correlation coefficient by classifying it into three categories. The first category was 'strong correlation' if 'r' value was greater than 0.8, the second category, 'good correlation', if 'r' value falls between 0.5 and 0.8 and the third category was 'moderate correlation', if 'r' value falls between 0.3 and 0.5 between the components of the study. In aggregate, the correlation between the components of financial inclusion and inclusive growth and the component of the socio-economic impact of availing formal financial services and inclusive growth was measured as 'good' and 'strong' correlation ('r=0.801' and '0.660' p <.01) respectively, statistically significant.

## 12. Regression Analysis

To evaluate the impact of availing formal financial services on the outcome of inclusive growth multiple regression analysis has been carried out utilizing 'inclusive growth' as criterion and 'financial inclusion' and 'impact of availing financial services' as predictors. The analysis was statistically significant (t = 30.044 & 26.257, p <.01), indicating that at the aggregate level, the components of 'financial inclusion' (nature of financial inclusion + extent of financial inclusion)  and the 'impact of availing financial services' (material changes + cognitive changes + perceptional changes + relational changes after availing financial services) had a significant impact on the component of 'inclusive growth' confirmed by the values of Standard Beta (β= 0.572 & 0.500) respectively.  This accounted for 0.622 of the variability as indexed by the R² statistic. The regression equation for predicting inclusive growth from financial inclusion and impact of access to finance was found to be Y = 13.335 + 0.774 × X1 + 0.848 × X2.

## 13. Policy Implications and Conclusions

 Based on observations, analysis, and findings, the paper draws the following conclusions:

(i) There is a disparity in financial inclusion at the macro level of the Indian economy and across the revenue divisions of Karnataka.  Even after initiatives are taken towards financial inclusion, at a macro level a vast segment of farmer households, poor and marginal sections of the society do not access any financial services and there is wide disparity throughout the economy. The disparities across the revenue divisions within the state, among the households belonging to vulnerable sections, are confirmed by the present study.

(ii) There is a positive correlation between availing of banking services and the socio-economic status of the households belonging to the vulnerable sections. Therefore, financial inclusion of the households belonging to vulnerable sections may be considered as an important factor in the improvement of the socio-economic status of the households.

(iii) The positive correlation between financial inclusion and inclusive growth may be considered as a factor to highlight the role of innovative, low-cost formal banking products in the process of inclusive growth.

(iv) Along with the basic banking products, households belonging to the vulnerable sections accessed human development packages, risk mitigation packages, and productivity enhancement packages. This phenomenon indicates the role of welfare packages in improving the absorptive capacity of vulnerable sections which takes them out of the vicious circle of financial exclusion and makes them more bankable. Therefore, households belonging to the vulnerable sections of the economy would be benefited from increased financial inclusion when it is promoted in a wider context of economic inclusion along with the welfare packages.

(v) The formal financial network could unleash the creative capacities of the households of vulnerable sections driven by income and consumption growth of a vast segment of the population belonging to vulnerable sections.

(vi) The formal banking system could lead to the betterment of the financial conditions and standard of living of households belonging to vulnerable sections empowering them to create assets, income, and emergency funds to meet unforeseen conditions and economic crisis. It also benefits by way of leakage-proof transfer, welfare benefits to the downtrodden sections of the population.

(vii) Financial inclusion broadens the banking network with a larger number of transactions leading to new opportunities to earn livelihood in the society which would lead to the inclusive growth of the economy.

(viii) Given the existence of financial exclusion in the context of the Indian economy, the present study rationalizes the focus of public policies on financial inclusion.

(ix) By confirming the correlation between financial inclusion and its impact, leading to positive changes in the socio-economic status of households in the survey area, the present study underscores the need of putting financial sector reforms that promote financial inclusion at the center of the development agenda.

Based on this conclusion, the study proposes a model to make the financial system more inclusive, which will contribute to the pro-poor growth in the economy. The model is explained in three stages (Figure 9)

**Stage I: Making Households Bankable by Improving the Absorptive Capacity.**

I stage of the model suggests methods to make households belonging to vulnerable sections bankable, which improves their absorptive capacity, and households come out of the vicious circle of financial exclusion. In this stage, human development packages on primary education, primary health, vocational training, etc., along with risk mitigation packages and productivity enhancement packages will be introduced. Government-sponsored campaigns through media, banking networks, NGO and SHG campaigns in strengthening the financial literacy and primary knowledge of the households in accessing formal financial services will play a deciding role at this stage. Local organizations and members join the stream to remove the apprehensions that the customers of vulnerable sections have towards the banking sector. Well-informed local volunteers will help the households to strengthen their position. The entire process of the stage first marks the beginning of inclusive growth in a true sense.

**Stage II: Providing Basic Banking Facilities to BPL Households**

In the second stage, the formal baking network will adopt a considerate approach towards new customers of the vulnerable sections in the form of no-frill accounts, facility of withdrawals, and fewer restrictions on such accounts. Facilities like e-transfer of benefits, financial advice to make use of the available financial products effectively, and delivery of all the products at one point will be provided, which will increase the confidence of households in accessing formal banking network. This will also improve the business opportunity of the banking network which will result in low-cost banking.

**Stage III: Innovative Strategies**

The third stage is supported by innovative strategies, would improve the reach of banking facilities. This is a phase of infrastructure sharing amongst banks and other organizations. The resulting low-cost benefit in this stage can be transferred to the households belonging to the vulnerable sections. To further strengthen the formal banking network, banks can open small extension counters at the organizations providing public utility services such as local schools, primary health care centers, village Panchayaths, farmers` centers, railway stations, bus stops, etc., along with the technological innovations.

In this way, the suggested model results in the improved absorptive capacity of households of vulnerable sections making them more bankable. Financial inclusion that has taken place, thus, broadens the access to formal finance and paves the way for the larger number of transactions, which will result in formalized and systematic banking; rapid economic growth; new opportunities for production and employment in the society, leading to inclusive growth.

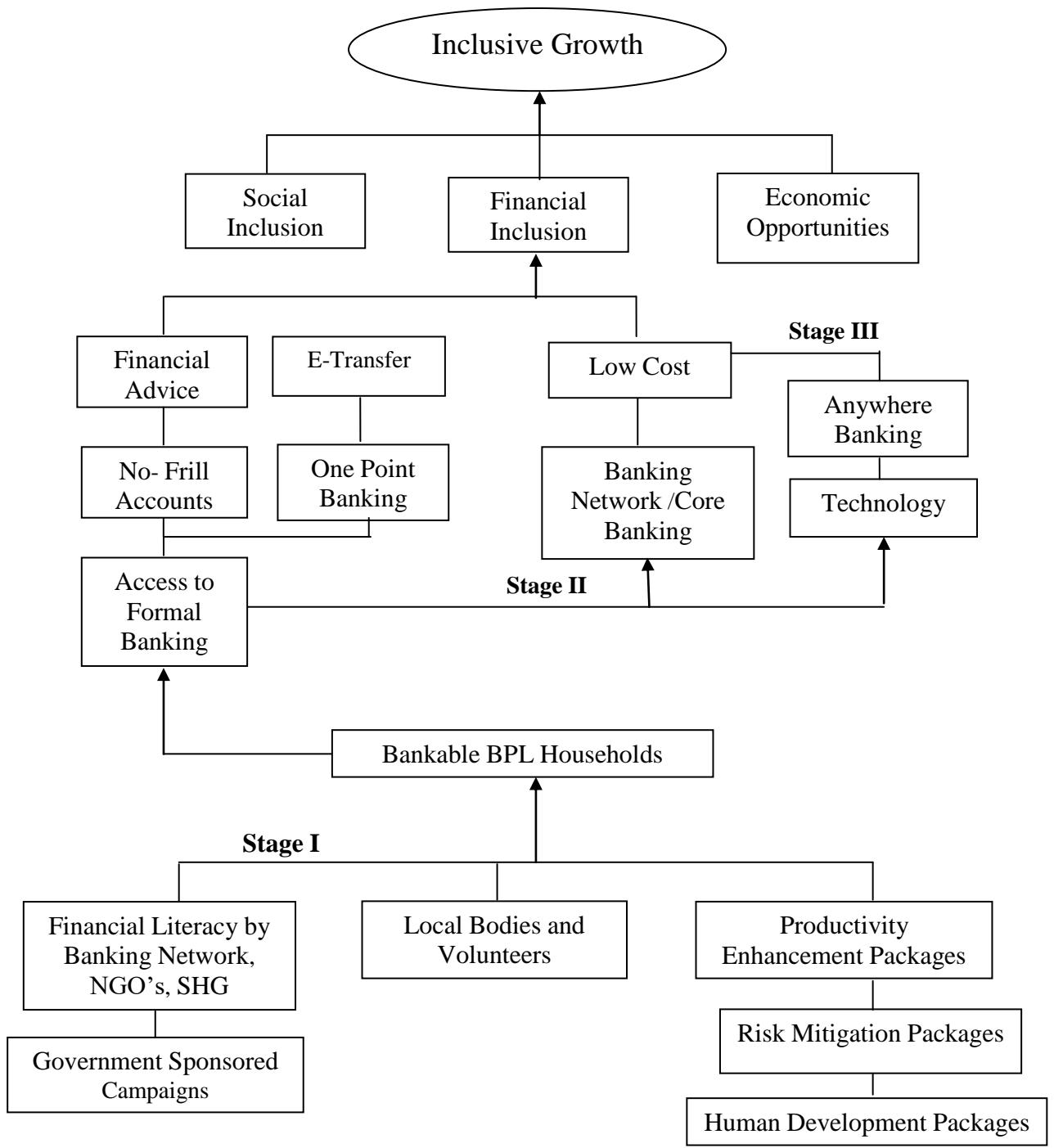

**Figure 9. Suggested Model on Financial Inclusion and Inclusive Growth**

## 14. Limitations of the Study

i) The foremost limitation of the study is that the analysis is restricted to the primary data gathered exclusively from the rural and urban taluks of the sample area.

ii) The basic assumption of the study is that other things being constant, access to formal banking network and availing of formal financial services lead to positive changes in the socio-economic conditions of the households, resulting in the 'inclusive growth of the economy. But the in-ground reality, there may be other factors contributing to the all-round development of the economy.

iii) A few more rural and urban villages could have been added to the list of sample areas to study the nature, extent, and impact of financial inclusion on the households for a wider range of geographical areas. Due to time and other resource constraints, such an analysis is not attempted in the present study. Given the small sample size, the data and findings from this research cannot be considered conclusive.

iv) The impact of access to banking services is measured by using the cross-sectional data collected by a snapshot method. A longitudinal study and panel data would have given robust information and results in this direction.

## References

Aghion, Philippe, Bolton, & Patrick. (1997). A Theory of Trickle-Down Growth and Development. *The Review of Economic Studies*, *64* (2),151-172.

Ahluwalia, M. S. (2011). Prospects and Policy Challenges in the Twelfth Plan. *Economic and Political Weekly*, *46* (21),88-105.

Batnagar, S. D. (2007). Emerging Knowledge-based innovations with special reference to Micro Credit. *The Journal of Indian Institute of Banking and Finance*, *78*,50-58.

Beck, Thorsten, Demirguc-Kunt, Asli, Honahan, & Patrick. (2009). Access to Financial Services: Measurement, Impact, and Policies. *The World Bank Research Observer,24,*1.Retrieved July 16, 2010, from https://openknowledge.worldbank.org/bitstream/handle/10986/4426/wbro_24_1_119.pdf?sequence=1.

Beck, Thorsten, Demirguc-Kunt, Asli, Peria, & Martinez, S. M. (2005). Reaching Out: Access to and Use of Banking Services Across Countries (World Bank Policy Research Working Paper No. 3754*)*. World Bank, Washington D.C.

Bell, C. (1990). Interactions between Institutional and Informal Credit Agencies in Rural India. *The World Bank Economic Review*, *4*(3), 297-327.

Chakrabarty, C. K. (2009). A Presentation on financial inclusion at 20th SKOCH Summit. Retrieved April 20, 2010, from www.rbi.org.in.

Collins, D., Morduch, J., Rutherford, S., & Ruthven, O. (2009). *Portfolios of the Poor: How the World's Poor Live on $2 a Day*. *Princeton,* NJ: Princeton University Press.

Leeladhar, V. (2005).Taking Banking Services to the Common Man: Financial Inclusion, *CommemorativeLecture*, Ernakalulam: Fedbank HormisMemorial Foundation.

Levine, & Ross. (1997). Financial Development and Economic Growth: Views and Agenda. *Journal of Economic Literature*,*35*,688-726.

Mohan, R. (2006). Economic Growth, Financial deepening, and Financial Inclusion. *Annual Bankers Conference.* Hyderabad.

Muhammad, Y. (1998). *Banker to the Poor: Micro-lending and the Battle against World Poverty*. New York: Public Affairs, Perseus Book Group.

Muhammad, Y., & Weber, K. (2007). *Creating a World without Poverty: Social Business and the Future of Capitalism.* New York: Public Affairs, Perseus Book Group.

NABARD. (2008). *NABCONS Consultancy Report on the Financial Inclusion survey of the 7 Districts of Karnataka.* Bangalore: NABARD.

National Bank for Agriculture and Rural Development. (2011). *Status of Microfinance in India 2008 to 2011.* Mumbai: NABARD.

Reserve Bank of India. (2008). *Report of the Committee on Financial Inclusion, GOI*. Retrieved September 15,2008, https://www.nabard.org/pdf/Full_Report.pdf.

Sarma, M. (2008). Index of Financial Inclusion (Working Paper, No.215*).*Indian Council for Research on International Economic Relations, New Delhi.

Serrao, M. V., Sequeira, A. H. (2012). Role of Financial Inclusion in Rural Transformation-A Need for Multidimensional Approach. In K.V.M. Varambally (Ed.), *Rural Transformation and Development Perspectives- Managerial Challenges* (pp 85-92). New Delhi: Excel India Publishers.

Serrao, M. V., Sequeira, A. H., Hans, V. B. (2012). Designing a Methodology to Investigate Accessibility and Impact of Financial Inclusion. In K. Gangadharan (Ed.), *Financial Inclusion and Inclusive Growth, Scope and Dimensions* (pp 180-199). New Delhi: Reference Press.

Swain, B., & Ranjula. (2002). Credit Rationing in Rural India. *Journal of Economic Development*, *27* (2), 1-19.

Thiel, M. (2001). Finance and Economic Growth – A Review of Theory and the available Evidence ( Economic Paper No 158). European Commission, Directorate-General for Economic and Financial Affairs, Brussels.

Thorat, U., (2006a). Financial Inclusion and Millennium Development Goals. *Reserve Bank of India Bulletin*, 239-243.

United Nations. (2006). *Building Inclusive Financial Sectors for Development*, Blue Book. Department of Economic and Social Affairs.Newyork: UN.

World Bank. (2008). Finance for All?: Policies and Pitfalls in Expanding Access (World Bank Policy Research Paper), World Bank, Washington D.C.

World Bank. (2009). *Banking the Poor: Measuring Banking Access in 54 Economies.* World Bank, Washington DC.